\documentclass[twocolumn,nofootinbib,superscriptaddress,showpacs]{revtex4}
\usepackage{amsmath,amssymb,amsbsy}
\usepackage{graphicx,subfigure}
\usepackage{comment}

\begin{document}

\unitlength=1mm

\def\a{{\alpha}}
\def\b{{\beta}}
\def\d{{\delta}}
\def\D{{\Delta}}
\def\e{{\epsilon}}
\def\g{{\gamma}}
\def\G{{\Gamma}}
\def\k{{\kappa}}
\def\l{{\lambda}}
\def\L{{\Lambda}}
\def\m{{\mu}}
\def\n{{\nu}}
\def\w{{\omega}}
\def\O{{\Omega}}
\def\S{{\Sigma}}
\def\s{{\sigma}}
\def\t{{\tau}}
\def\th{{\theta}}
\def\x{{\xi}}

\def\ol#1{{\overline{#1}}}

\def\Dslash{D\hskip-0.65em /}
\def\dslash{{\partial\hskip-0.5em /}}
\def\vslash{{\rlap \slash v}}
\def\qbar{{\overline q}}

\def\CPT{{$\chi$PT}}
\def\QCPT{{Q$\chi$PT}}
\def\PQCPT{{PQ$\chi$PT}}
\def\tr{\text{tr}}
\def\str{\text{str}}
\def\diag{\text{diag}}
\def\order{{\mathcal O}}
\def\vit{{\it v}}
\def\vD{\vit\cdot D}
\def\am{\alpha_M}
\def\bm{\beta_M}
\def\gm{\gamma_M}
\def\smb{\sigma_M}
\def\smt{\overline{\sigma}_M}
\def\tb{{\tilde b}}

\def\mc#1{{\mathcal #1}}

\def\Bbar{\overline{B}}
\def\Tbar{\overline{T}}
\def\cBbar{\overline{\cal B}}
\def\cTbar{\overline{\cal T}}
\def\pq{(PQ)}

\def\eqref#1{{(\ref{#1})}}

\preprint{UMD-40762-432}

\title{\textbf{$\pi\pi$ Scattering in Twisted Mass Chiral Perturbation Theory}}

\author{Michael I. Buchoff}
\email[]{mbuchoff@umd.edu}
\affiliation{Maryland Center for Fundamental Physics, \\
Department of Physics, University of Maryland,
	College Park, MD 20742-4111}

\author{Jiunn-Wei Chen}
\email[]{jwc@phys.ntu.edu.tw} 
\affiliation{Department of Physics and Center for Theoretical Sciences, National Taiwan University, Taipei 10617, Taiwan}

\author{Andr\'e Walker-Loud} 
\email[]{walkloud@wm.edu}
\affiliation{Department of Physics, College of William and Mary,
	P.O. Box 8795,
	Williamsburg, VA 23187-8795, USA}

\begin{abstract}
In this report we describe both $I=2$ and $I=0$ $\pi\pi$ scattering for twisted mass lattice QCD utilizing twisted mass chiral perturbation theory at next-to-leading order.  Focusing on the lattice spacing ($b$) corrections, we demonstrate that in the exotic $I=2, I_3=\pm2$ channels ($\pi^\pm\pi^\pm$), the leading scaling violations of $\pi\pi$ scattering at maximal twist begin at $\mathcal{O}(m_\pi^2 b^2)$.  This is not the case in any other isospin channel, for which the scaling violations at maximal twist begin at $\mathcal{O}(b^2)$.  Furthermore, we demonstrate the existence of a mixing between the $I=2, I_3 = 0$ and $I=0$ scattering channels due to the breaking of isospin symmetry by the twisted mass term.  The mixing term, although formally next-to-leading order, is relatively large, thus necessitating the use of a coupled channel analysis.  We argue that this mixing likely renders the computation of the $I=0$ channel impractical with twisted mass lattice QCD.
\end{abstract}

\maketitle

\section{Introduction}
The last few years have seen a growth in the computation of hadron interactions with lattice QCD~\cite{Beane:2008dv}, with dynamical calculations of two-meson systems~\cite{Yamazaki:2004qb,Beane:2005rj,Beane:2007xs,Beane:2006gj,Beane:2007uh}, two-baryon systems~\cite{Beane:2006mx,Beane:2006gf} and systems of up to 12 pions~\cite{Beane:2007es,Detmold:2008fn} and kaons~\cite{Detmold:2008yn}.  Further, lattice field theory methods are now being applied to the low-energy effective field theory of multinucleon interactions~\cite{Seki:2005ns,Chen:2003vy,Chen:2004rq,Lee:2004si,Lee:2004qd,Borasoy:2006qn,Borasoy:2007vi,Borasoy:2007vk}, for which there exists a nice review~\cite{Lee:2008fa}.  The dynamical lattice QCD calculations of hadron interactions to date, have either been performed with Wilson fermions or a mixed lattice action~\cite{Renner:2004ck,Edwards:2005kw} of domain-wall valence fermions~\cite{Kaplan:1992bt,Shamir:1993zy,Furman:1994ky} and the Asqtad improved~\cite{Orginos:1998ue,Orginos:1999cr} rooted staggered MILC configurations~\cite{Bernard:2001av,Aubin:2004wf}.  Twisted mass lattice QCD~\cite{Frezzotti:1999vv,Frezzotti:2000nk} has recently emerged as a viable fermion discretization method for computing gauge configurations with two flavors of light quarks, ($up$ and $down$) ~\cite{Boucaud:2007uk,Blossier:2007vv,Boucaud:2008xu} and hopeful prospects of $2+1+1$ ($up$, $down$, $strange$ and $charm$) flavors of dynamical sea fermions~\cite{Chiarappa:2006ae} in the chiral regime.  It is therefore only a matter of time before hadron interactions will be computed with the twisted mass fermion discretization method.

The first multihadron system to be explored with twisted mass lattice QCD will most likely be that of two pions.
The two-pion system is numerically the simplest as well as theoretically the best understood.  In fact, the scattering of two pions at low energies was uniquely predicted at leading order (LO) in chiral perturbation theory ($\chi$PT) by Weinberg in 1966~\cite{Weinberg:1966kf}.  The subleading orders in the chiral expansion give rise to perturbative corrections to the LO predictions and have been worked out to one-loop, or next-to-leading order (NLO) by Gasser and Leutwyler~\cite{Gasser:1983yg} and also to two-loops, or next-to-next-to-leading order (NNLO)~\cite{Knecht:1995tr,Bijnens:1995yn,Bijnens:1997vq}.  Each new order introduces operators with coefficients not constrained by chiral symmetry, known as low-energy constants (LECs).  To have predictive power, these LECs must be determined either by comparison with experiment or lattice QCD calculational results.  A comparison with lattice QCD can introduce additional complications as the calculations are performed at finite lattice spacing in a finite volume.  Modifications to the infrared and ultraviolet behavior of the theory can be incorporated into chiral perturbation theory.  For sufficiently large but finite lattice volumes, the operator structure and power counting of the effective theory remain valid with exponentially small corrections to matrix elements~\cite{Gasser:1987zq}.  Lattice discretization effects can also be incorporated into the chiral Lagrangian through a two-step process first detailed in Ref.~\cite{Sharpe:1998xm}.  One first constructs the effective continuum Symanzik Lagrangian~\cite{Symanzik:1983dc,Symanzik:1983gh} for a given lattice action.  One then builds the low energy chiral Lagrangian from the Symanzik theory, giving rise to new unphysical operators with their own LECs.  These new operators capture the discretization effects for a given lattice action.

In this report, we briefly review the construction of the twisted mass chiral Lagrangian in Sec.~\ref{sec:tmLag}.
We then determine the lattice spacing $(b)$ corrections to low-energy $\pi\pi$ scattering specific to the twisted mass lattice action.  We work through $\mc{O}(b^2)\sim\mc{O}(bm_\pi^2)$.

\section{Twisted mass lattice QCD and the continuum effective action\label{sec:tmLag}}

The twisted mass chiral Lagrangian was determined previously in Refs.~\cite{Munster:2003ba,Scorzato:2004da,Sharpe:2004ps,Aoki:2004ta,Sharpe:2004ny}, and for baryons in Ref.~\cite{WalkerLoud:2005bt}.  In this report we focus on twisted mass lattice QCD with degenerate light flavors given by the lattice action%
%
\begin{align}
S =&\ \sum_{x} \bar{\psi}(x) \bigg[ \frac{1}{2} \sum_\nu \g_\nu (\nabla^*_\nu + \nabla_\nu)
	-\frac{r}{2} \sum_\nu \nabla^*_\nu \nabla_\nu
\nonumber\\&
	+m_0 + i\g_5 \t_3 \mu_0 \bigg] \psi(x)\, ,
\end{align}
where $\psi$ and $\bar{\psi}$ are the dimensionless lattice fermion fields, $\nabla_\nu(\nabla_\nu^*)$ are the covariant forward (backward) lattice derivatives in the $\nu$ direction, $m_0$ is the dimensionless bare quark mass and $\mu_0$ is the dimensionless bare twisted quark mass.  The fermion fields are flavor doublets, $\tau_3$ is the third Pauli-spin matrix and the bare mass term is implicitly accompanied by a flavor identity matrix.  Our twisted mass $\chi$PT analysis also holds for dynamical lattice calculations with $2+1+1$ flavors, the only difference being the numerical values of the LECs determined when fitting the extrapolation formula to the calculation results.

The continuum chiral Lagrangian, supplemented by discretization effects is determined with the two step procedure of Ref.~\cite{Sharpe:1998xm}.  This was done for the twisted mass lattice action in Ref.~\cite{Sharpe:2004ps}, to NLO in which a power counting $m_q \sim b \Lambda_{QCD}^2$ was used and which we shall adopt.  The resulting effective  Lagrangian is
\begin{align}\label{eq:Leff}
\mc{L}_{eff} =&\ \mc{L}_{glue}
	+\bar{q} (\Dslash + m + i \g_5 \t_3 \mu ) q
\nonumber\\&
	+c_{SW} b\ \bar{q}\, i \s_{\mu\nu} F_{\mu\nu} q
	+\mc{O}(b^2, bm_q, m_q^2)\, ,
\end{align}
where $\mc{L}_{glue}$ is the Yang-Mills Lagrangian.  The quark fields are an isodoublet, $q^T = (q_u, q_d)$ and the quark masses are given by
\begin{align}
	m &= Z_m (m_0 - m_c) / b\, ,
\nonumber\\
	\mu &= Z_\mu \mu_0 / b\, .
\end{align}
The symmetry properties of the twisted mass lattice action protect the twisted mass from additive mass renormalization.  With Eq.~\eqref{eq:Leff}, one can construct the two flavor chiral Lagrangian.  This is the Gasser-Leutwyler Lagrangian~\cite{Gasser:1983yg} supplemented by chiral and flavor symmetry breaking terms proportional to the lattice spacing.  The Lagrangian through NLO relevant to our work takes the form~\cite{Sharpe:2004ps,Sharpe:2004ny} (we use the normalization $f\sim 130$~MeV),
\begin{widetext}
\begin{align}\label{eq:Lag}
\mc{L}_\chi^{tw} =&\ \frac{f^2}{8} \tr (\partial_\mu \S \partial_\mu \S^\dagger )
	-\frac{f^2}{8}\tr ( \chi^{\prime \dagger} \S + \S^\dagger \chi^\prime )
	-\frac{l_1}{4} \tr (\partial_\mu \S \partial_\mu \S^\dagger )^2
	-\frac{l_2}{4} \tr (\partial_\mu \S \partial_\nu \S^\dagger ) 
		\tr (\partial_\mu \S \partial_\nu \S^\dagger )
\nonumber\\&
	-\frac{l_3+l_4}{16} \Big[ \tr ( \chi^{\prime \dagger}\S + \S^\dagger \chi^\prime) \Big]^2
	+\frac{l_4}{8} \tr (\partial_\mu \S \partial_\mu \S^\dagger )
		\tr ( \chi^{\prime \dagger}\S + \S^\dagger \chi^\prime)
\nonumber\\&
	+\tilde{W} \tr (\partial_\mu \S \partial_\mu \S^\dagger )
		\tr (\hat{A}^\dagger \S + \S^\dagger \hat{A} )
	-W \tr (\chi^{\prime \dagger}\S + \S^\dagger \chi^\prime)
		\tr (\hat{A}^\dagger \S + \S^\dagger \hat{A} )
	-W^\prime \Big[ \tr (\hat{A}^\dagger \S + \S^\dagger \hat{A} ) \Big]^2\, ,
\end{align}
\end{widetext}
where the LECs, $l_1$--$l_4$ are the $SU(2)$ Gasser-Leutwyler coefficients and the coefficients $\tilde{W}$, $W$ and $W^\prime$ are unphysical LECs arising from the explicit chiral symmetry breaking of the twisted mass lattice action.  The spurion fields are defined as
\begin{align}
	&\chi^\prime = 2B_0(m + i\t_3 \mu) + 2W_0 b \equiv \hat{m} + i\t_3 \hat{\mu} + \hat{b}&
\nonumber\\
	&\hat{A} = 2W_0 b \equiv \hat{b}\, .&
\end{align}
As discussed in Ref.~\cite{Sharpe:2004ny}, the vacuum of the theory as written is not aligned with the flavor identity but is given at LO by
\begin{equation}
\S_0 \equiv \langle 0 | \S | 0 \rangle
	= \frac{\hat{m} + \hat{b} + i\t_3 \hat{\mu}}{M^\prime}
	= \textrm{exp}(i \w_0 \t_3)\, ,
\end{equation}
with
\begin{equation}
	M^\prime = \sqrt{(\hat{m}+\hat{b})^2 +\hat{\mu}^2}\, .
\end{equation}
Therefore, to determine the Feynman rules which leave the interactions of the theory the most transparent, one expands the Lagrangian around the physical vacuum.    Extending this analysis to NLO, one finds the vacuum angle shifts to $\w = \w_0 + \e$ where one can determine $\e$ either by finding the minimum of the potential, as was done in Ref.~\cite{Sharpe:2004ny} or by requiring the single pion vertices to vanish,
\begin{equation}\label{eq:deltaw}
	\e(\w_0) = -\frac{32}{f^2}\hat{b} \sin \w_0 \left[ W + 2W^\prime \cos \w_0 \frac{\hat{b}}{M^\prime} \right]\, .
\end{equation}
One can expand about the physical vacuum by making the replacement
\begin{align}
	&\S = \xi_m\, \S_{ph}\, \xi_m,&
	&\textrm{with}&
	&\xi_m = \textrm{exp} (i \w \t_3 / 2),&
\end{align}
and
\begin{align}
&\S_{ph} = \textrm{exp} \left( \frac{2i \phi}{f} \right),& 
&\phi = \begin{pmatrix} \frac{\pi^0}{\sqrt{2}} & \pi^+ \\ \pi^- & -\frac{\pi^0}{\sqrt{2}}
	\end{pmatrix}\, .&
\end{align}
One then finds the Lagrangian is given by
\begin{widetext}
\begin{align}
\mc{L} =&\  \mc{L}_{cont.} +
	\tilde{W} \hat{b} \cos \w\, \tr ( \partial_\mu \S_{ph} \partial_\mu \S_{ph}^\dagger) 
		\tr (\S_{ph} + \S_{ph}^\dagger)
	-\hat{b} \cos \w \left( W M^\prime +W^\prime \hat{b} \cos \w \right)
		\left[ \tr (\S_{ph} + \S_{ph}^\dagger) \right]^2
\nonumber\\&
	+\tilde{W} \hat{b}\sin \w\, \tr ( \partial_\mu \S_{ph} \partial_\mu \S_{ph}^\dagger) 
		\tr ( i\t_3 (\S_{ph} - \S_{ph}^\dagger ) )
	-W^\prime \hat{b}^2 \sin^2 \w\, 
		\left[ \tr ( i\t_3 (\S_{ph} - \S_{ph}^\dagger ) ) \right]^2
\nonumber\\&
	-\tr ( i\t_3 (\S_{ph} - \S_{ph}^\dagger ) ) \left[
		\e(\w)\, \frac{M^\prime f^2}{8}
		+\hat{b} \sin \w\, \left( W M^\prime + 2 W^\prime \hat{b} \cos \w\, \right)
			\tr (\S_{ph} + \S_{ph}^\dagger) \right]\, ,
\end{align}
\end{widetext}
where $\mc{L}_{cont.}$ is the continuum $SU(2)$ chiral Lagrangian to NLO.  
Of particular interest to us are the new two, three and four pion interactions which result from the discretization errors in the twisted mass Lagrangian.
We find, in agreement with Ref.~\cite{Sharpe:2004ny}
\begin{equation}
\mc{L} = \mc{L}_{cont.} + \D\mc{L}_{2\phi} + \D\mc{L}_{3\phi} + \D\mc{L}_{4\phi}\, ,
\end{equation}
where
\begin{align}\label{eq:L2phi}
\D\mc{L}_{2\phi} = &\ 
	\cos \w \frac{16\tilde{W}\hat{b}}{f^2} \tr (\partial_\mu \phi \partial_\mu \phi )
	+\frac{1}{2}\D M^\prime(\w)\, \tr( \phi^2 ) 
\nonumber\\&
	+ \frac{1}{2} \D M^\prime_0(\w)\, \left[ \tr \left( \frac{\t_3 \phi}{\sqrt{2}} \right) \right]^2\, ,
\end{align} 
\begin{align}\label{eq:L3phi}
\D\mc{L}_{3\phi} =&
	-\sin \w\, \frac{16 \tilde{W} \hat{b}}{f^3} \tr(\t_3 \phi) \tr( \partial_\mu \phi \partial_\mu \phi)
\nonumber\\&
	+\frac{\e(\w) M^\prime}{2f} \tr( \t_3 \phi) \tr(\phi^2)\, ,
\end{align}
\begin{align}\label{eq:L4phi}
\D\mc{L}_{4\phi} =&
	-\frac{\D M^\prime(\w)}{3 f^2} [\tr ( \phi^2 ) ]^2
	-\frac{\D M^\prime_0(\w)}{3f^2} [\tr (\frac{\t_3 \phi}{\sqrt{2}} )]^2 \tr (\phi^2 )
\nonumber\\&
	+\cos \w\, \frac{32 \tilde{W} \hat{b}}{3f^4} \tr ( \phi\, \partial_\mu \phi [ \phi, \partial_\mu \phi] )
\nonumber\\&
	-\cos \w\, \frac{16 \tilde{W} \hat{b}}{f^4} \tr(\partial_\mu \phi \partial_\mu \phi ) \tr ( \phi^2)\, ,
\end{align}
and the mass corrections are given by
\begin{align}
	&\D M^\prime(\w) = \cos \w \frac{64 \hat{b}}{f^2} \left( W M^\prime + \cos \w\, W^\prime \hat{b} \right),&
\nonumber\\ \label{eq:Mprime0}
	&\D M^\prime_0(\w) = -\sin^2 \w \frac{64 W^\prime \hat{b}^2}{f^2}\, .&
\end{align}
From this Lagrangian, one can determine the pion masses, decay constants and wave-function corrections.  One finds the masses are (using the modified \textit{dimensional regularization} of Ref.~\cite{Gasser:1983yg})
\begin{align}
m_{\pi^\pm}^2 =&\ M^\prime \left[
	1 + \frac{M^\prime}{(4\pi f)^2}\ln \left( \frac{M^\prime}{\mu^2} \right) + l_3^r(\mu)\frac{4 M^\prime}{f^2} \right]
\nonumber\\&
	+\D M^\prime(\w) - \cos\w \frac{32\tilde{W}\hat{b}M^\prime}{f^2}\, ,
\\ \label{eq:mpi0}
m_{\pi^0}^2 =&\ m_{\pi^\pm}^2 + \D M^\prime_0(\w)\, ,
\end{align}
the decay constants are%
\footnote{There is an exact Ward identity one can exploit to compute the charged pion decay constant and avoid issues of the axial current renormalization discussed for example in Ref.~\cite{Aoki:2007es}.} 
\begin{align}
f_{\pi} =&\ f \bigg[ 
	1 - \frac{2M^\prime}{(4\pi f)^2} \ln \left(\frac{M^\prime}{\mu^2} \right)
	+l_4^r(\mu)\frac{2 M^\prime}{f^2} 
\nonumber\\&\quad
	+\cos\w \frac{16\tilde{W}\hat{b}}{f^2} \bigg]\, ,
\end{align}
and the wave-function correction is
\begin{align}
\d\mc{Z}_\pi = &\ \frac{4M^\prime}{3(4\pi f)^2} \ln \left( \frac{M^\prime}{\mu^2} \right)
	-l_4^r(\mu) \frac{4M^\prime}{f^2}
	-\cos\w \frac{32\tilde{W}\hat{b}}{f^2}
\end{align}
These expressions will be needed to express the scattering in terms of the lattice-physical parameters (by lattice-physical, we mean the renormalized mass and decay constant as measured from the correlation functions, and not extrapolated to the continuum or infinite volume limit).  As we discuss in the next section, these interactions lead to three types of new contributions to $\pi\pi$ scattering states: there are discretization corrections to the scattering parameters, the scattering lengths, effective ranges, etc., which appear in a mild manner as those from the Wilson chiral Lagrangian~\cite{Buchoff:2008ve}.  There are corrections which can potentially significantly modify the chiral behavior, arising from the three-pion interactions, and there are new corrections which mix different scattering channels, for example, the $I=2, I_3=0$ and the $I=0$ scattering states.

%
\section{$\pi\pi$ Scattering in Twisted Mass $\chi$PT}
In this section we calculate corrections to the two-pion scattering channels.  We begin with the maximally stretched $I=2$ states, which have the simplest corrections.

\subsection{$I=2, I_3=\pm 2$ Channels}
There are two types of discretization corrections which modify the $I=2, I_3=\pm2$ scattering, those which are similar to the corrections for the Wilson lattice action~\cite{Buchoff:2008ve,Aoki:2008gy} and those which arise from the three-pion interactions, Eq.~\eqref{eq:L3phi} and give rise to new Feynman diagrams.  We will express the scattering parameters in terms of the lattice-physical pion mass and decay constant.  As was shown in detail, this has dramatic consequences on the formula for the scattering parameters in both partially quenched and mixed action $\chi$PT~\cite{Chen:2005ab,Chen:2006wf,Chen:2007ug}, such that the extrapolation formulae were free of unphysical counterterms through NLO.  There is a second benefit to expressing the scattering parameters in lattice-physical parameters.  This allows one to perform a chiral extrapolation in terms of the ratio $m_\pi / f_\pi$, and thus avoid the need for scale setting.  This was crucial in allowing the NPLQCD Collaboration to make a precision prediction of the $I=2$ scattering length~\cite{Beane:2005rj,Beane:2007xs}.

\begin{figure}[t]
\begin{tabular}{ccc}
\includegraphics[width=0.1\textwidth]{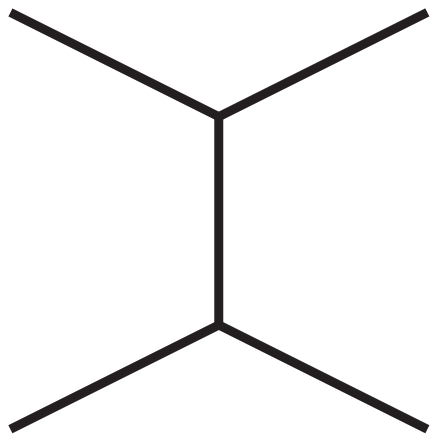}
&\phantom{space}&
\includegraphics[width=0.1\textwidth]{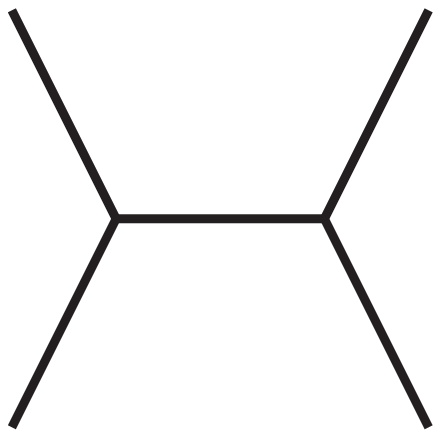}
\\
$(a)$ &&$(b)$
\end{tabular}
\caption{\label{fig:3phi_graphs} New unphysical graphs from twisted mass interactions in the $t$($u$)-channel $(a)$ and $s$-channel $(b)$.  Fig.~$(b)$ can only contribute to $I_3=0$ scattering.}
\end{figure}

The simple corrections to the scattering amplitude are determined from $\D\mc{L}_{4\phi}$, Eq.~\eqref{eq:L4phi}.  The three-pion interactions from Eq.~\eqref{eq:L3phi} lead to new topological graphs in the scattering amplitude, which we depict in Fig.~\ref{fig:3phi_graphs}.  The $I=2, I_3=\pm2$ scattering channels receive corrections from Fig.~\ref{fig:3phi_graphs}$(a)$ and its $u$-channel counterpart.
The internal propagating pion is a $\pi^0$, which, for present lattice actions, is known to be lighter than the charged pions in dynamical twisted mass lattice calculations with degenerate light quark masses~\cite{Boucaud:2007uk,Boucaud:2008xu}.

Putting all the corrections together, one finds the scattering amplitude, which we express in Minkowski-space, is given by
\begin{equation}
\mc{T}^{2,\pm2} = \mc{T}^{2,\pm2}_{cont.} + \D\mc{T}^{2,\pm2}\, ,
\end{equation}
where the discretization corrections are 
\begin{align}
\D \mc{T}^{2,\pm2}(\w) &= 4\frac{\D M^\prime(\w)}{f_\pi^2}
	+\cos\w \frac{64\tilde{W} \hat{b}}{f_\pi^4}\, s
\nonumber\\&
	+\frac{2\e^2(\w)m_\pi^4}{f_\pi^2} \left( \frac{1}{m_{\pi^0}^2 - t} + \frac{1}{m_{\pi^0}^2 - u} \right)\, .
\end{align}
The first two terms arise from Eq.~\eqref{eq:L4phi} as well as from the conversion to the lattice-physical parameters.  The second two terms arise from Fig.~\ref{fig:3phi_graphs}$(a)$.  These terms are formally NNLO.  However, depending upon the precision with which the twist angle is tuned, these terms may become large and require promotion to lower order.  Expanding the NLO contribution to the twist angle, Eq.~\eqref{eq:deltaw}, one finds
\begin{equation}
\e^2(\w) m_\pi^4 = \left(\frac{64 W^\prime \hat{b}^2}{f_\pi^2}\right)^2 \sin^2 \w \cos^2 \w
	+\mc{O}(m_\pi^2)\, .
\end{equation}
We can then determine the corrections to the $I=2, I_3=\pm2$ scattering lengths, for which we find
\begin{align}
\D m_\pi a_{\pi\pi}^{I=2,\pm2}(\w) =&\
	\frac{\D M^\prime(\w)}{8\pi f_\pi^2}
	-\cos(\w)\frac{8 \tilde{W} \hat{b} m_\pi^2}{\pi f_\pi^4}
\nonumber\\& 
	+\frac{(32W^\prime)^2}{2\pi} \frac{\sin^2 \w \cos^2 \w}{m_{\pi^0}^2/f_\pi^2} \frac{\hat{b}^4}{f_\pi^8}\, .
\end{align}
The first observation we make is that at maximal twist, $\w=\pi/2$, these leading discretization errors exactly cancel through NLO (this is true of the corrections to the scattering amplitude and not just the scattering length)%
\footnote{We have assumed that a suitable definition of the maximal twist angle has been used in the numerical lattice computations such that one is not restricted to the regime $m_q >> b^2 \Lambda_{QCD}^3$, but rather one is allowed $m_q \gtrsim b \L_{QCD}^2$~\cite{Frezzotti:2003ni,Aoki:2004ta,Sharpe:2004ny,Sharpe:2005rq}.} 
\begin{equation}
	\D m_\pi a_{\pi\pi}^{I=2,\pm2}(\pi/2) = 0\, .
\end{equation}
This is independent of the use of lattice-physical parameters, and holds also for the scattering length expressed in bare parameters, or any combination of bare and physical.  At zero twist, $\w=0$, our expressions reduce to those of Ref.~\cite{Buchoff:2008ve}.  Converting $f_\pi \rightarrow f$, our answer agrees with that in Ref.~\cite{Aoki:2008gy}.  The scattering length at maximal twist is simply given by the continuum formula
\begin{multline}
m_\pi a_{\pi\pi}^{I=2} = -2\pi \left(\frac{m_\pi}{4\pi f_\pi}\right)^2 \bigg\{
	1 + \left(\frac{m_\pi}{4\pi f_\pi}\right)^2
\\
	\times \left[ 
		3 \ln \left( \frac{m_\pi^2}{\mu^2} \right)
		-1 -l_{\pi\pi}^{I=2}(\mu) \right]
	\bigg\}\, ,
\end{multline}
where the combination of Gasser-Leutwyler coefficients is~\cite{Bijnens:1995yn,Bijnens:1997vq}
\begin{equation}
	l_{\pi\pi}^{I=2} = 4(4\pi)^2 (4l_1^r +4l_2^r + l_3^r - l_4^r )\, .
\end{equation}
Furthermore, the discretization errors only enter at tree level at this order (when the expression is expressed in lattice-physical parameters), and thus at arbitrary twist, the exponentially suppressed finite volume corrections to L\"{u}scher's method are also given by those determined in continuum finite volume $\chi$PT~\cite{Bedaque:2006yi}.

Returning to the new graphs arising from the three-pion interactions, we can estimate the size of the corrections to the scattering amplitude using the known mass splitting between the charged and neutral pions~\cite{Boucaud:2007uk,Boucaud:2008xu}.  Estimating the splitting with the leading correction, Eq.~\eqref{eq:mpi0}, and solving for $W^\prime$ from Eq.~\eqref{eq:Mprime0}, we can estimate the corrections to the $I=2, I_3=\pm2$ scattering length near maximal twist.  As a ratio to the LO prediction for the scattering length, one finds
\begin{equation}
\frac{\D m_\pi a_{\pi\pi}^{I=2,\pm2}(\w)}{m_\pi^2 / 8\pi f_\pi^2} \simeq
	\frac{\cot^2\w \left(\D M^\prime_0(\w) / m_\pi^2 \right)^2}{1 + \D M^\prime_0(\w) / m_\pi^2}\, .
\end{equation}
At the lightest mass point calculated in Refs.~\cite{Boucaud:2007uk,Boucaud:2008xu}, which corresponds to $m_\pi \simeq 300$~MeV, the pion mass splitting is
\begin{equation}\label{eq:pionSplitting}
	\frac{\D M^\prime_0(\w\sim \pi/2)}{m_\pi^2} \simeq - 0.33\, ,
\end{equation}
and therefore one must have $\cot\w \geq 0.3$ for this term to make more than a 1\% correction.  Therefore, for current twisted mass lattice calculations, corrections to the $I=2, I_3=\pm2$ scattering length (and other parameters) should be negligible provided higher order corrections are as small as expected.

\subsection{$I_3=0$ scattering channels\label{sec:I3=0}}
There are several features which make scattering in the $I_3=0$ channels more complicated than in the $I=2,I_3=\pm2$ channels, most of which stem from the fact that the twisted mass lattice action explicitly breaks the full $SU(2)$ symmetry down to $U(1)$, the conserved $I_3$ symmetry.  The first technical complication is not specific to twisted mass calculations, but is simply the need to compute quark disconnected diagrams.  The second complication stems from the mass splitting of the charged and neutral pions.  Generally, one determines the scattering phase shift for two particles with the L\"{u}scher method~\cite{Maiani:1990ca,Hamber:1983vu,Luscher:1986pf,Luscher:1990ux}, by determining the interaction energy
\begin{equation}
	\D E_{\pi\pi} = 2\sqrt{p^2 + m_\pi^2} - 2m_\pi\, .
\end{equation}
In the isospin limit, the $|2,0\rangle$ and $|0,0\rangle$ states (in the $|I,I_3\rangle$ basis) are given by
\begin{align}\label{eq:pipiStates}
|2,0\rangle &= 
	\frac{1}{\sqrt{6}}\left( |\pi^+ \pi^-\rangle + |\pi^- \pi^+ \rangle - 2|\pi^0\pi^0\rangle \right)\, ,
\nonumber\\
|0,0\rangle &= 
	\frac{1}{\sqrt{3}}\left( |\pi^+ \pi^-\rangle + |\pi^- \pi^+ \rangle + |\pi^0\pi^0\rangle \right)\, .
\end{align}
However, given the relatively large mass splitting in current twisted mass lattice calculations, Eq.~\eqref{eq:pionSplitting}, the propagating eigenstates will be arbitrarily shifted from the physical states, perhaps shifting nearly to the $\{|\pi^\pm \pi^\mp\rangle, |\pi^0 \pi^0\rangle \}$ basis.  This would have to be disentangled numerically.  Even ignoring this issue, which we deem the most serious, and working with the continuum $\{|2,0\rangle, |0,0\rangle \}$ basis, there is a mixing of these states, which first appears at NLO as the second operator in Eq.~\eqref{eq:L4phi}.  Working with the states
\begin{equation}
| I,0 \rangle = 
	\begin{pmatrix}
	| 2, 0 \rangle \\
	| 0, 0 \rangle
	\end{pmatrix}
\end{equation}
one finds
\begin{equation}
\D \mc{T}^{2,0;0,0}_{\D\mc{L}_{4\phi}} = \frac{8\D M^\prime_0(\w)}{9f^2}
	\begin{pmatrix} 
		 4 & -\frac{7}{\sqrt{2}} \\
		-\frac{7}{\sqrt{2}} & 5
	\end{pmatrix}\, .
\end{equation}
Given the correction $\D M^\prime_0(\w)$, one sees this mixing is in fact maximal at maximal twist.  This is nominally a NLO effect, thus possibly leaving the states mostly unmixed.  However, a comparison of this term with the LO amplitude of the $I=2, I_3=\pm2$ scattering, one finds close to maximal twist
\begin{equation}
\frac{\D \mc{T}^{2,0;0,0}_{\D\mc{L}_{4\phi}} / (32\pi)}{-2\pi (m_\pi / 4\pi f_\pi)^2} 
	\simeq \frac{1}{m_\pi^2 / f_\pi^2}
		\begin{pmatrix} 
		 1.11 & -1.37 \\
		-1.37 & 1.38
		\end{pmatrix}\, .
\end{equation}
For $m_\pi / f_\pi =2$, all terms in this scattering matrix are approximately $1/3$ the size of the LO amplitude.  Since we now know that the physical NLO corrections to the $I=2$ scattering length for example, only provide a few percent deviation from the LO term~\cite{Beane:2005rj,Beane:2007xs,Chen:2005ab,Chen:2006wf}, we conclude that this NLO operator in fact provides a relatively large contribution to the scattering amplitude, and furthermore provides a large mixing term, and thus cannot be neglected.  This, combined with the problem we mentioned previously, means a coupled channel version of L\"{u}scher's method of determining the scattering parameters would be needed to explore the $I_3=0$ scattering channels with twisted mass lattice QCD.
\begin{figure}[t]
\begin{tabular}{c}
\includegraphics[width=0.45\textwidth]{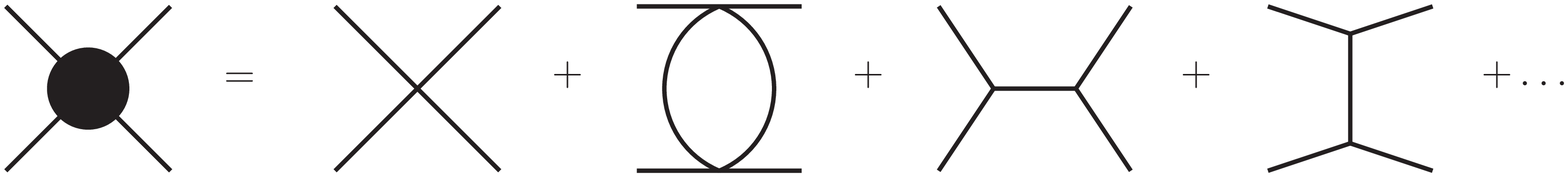}
\\
$(a)$ \\ \\
\includegraphics[width=0.45\textwidth]{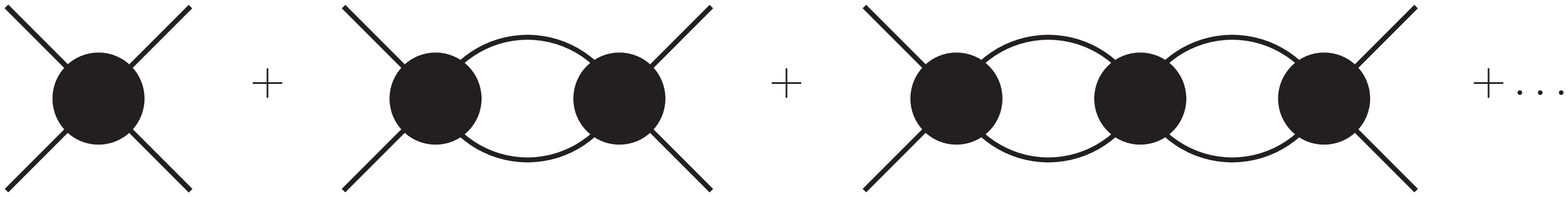}
\\
$(b)$
\end{tabular}
\caption{\label{fig:vertex} Modified four-point function, $(a)$ consisting of all off-shell graphs.  These vertices can then be iterated and summed $(b)$, to determine the $\pi\pi$ interactions.  This summation gives rise to the L\"{u}scher relation, valid below the inelastic threshold.}
\end{figure}

The $\D\mc{L}_{4\phi}$ Lagrangian is not the only source of mixing.  The three-pion interactions, depicted in Fig.~\ref{fig:3phi_graphs}, in the $s$, $t$ and $u$ channels, will also lead to a mixing of the $|2,0\rangle$ and $|0,0\rangle$ states, as one can check with an explicit calculation.  One may be concerned that the new $s$ channel graph will invalidate L\"{u}scher's method.  This is not the case however, as the internal pion propagator is always off-shell, and thus these diagrams do not contribute to the power-law volume dependence of the two-particle energy levels.  An alternative way to understand this is diagrammatically.  One can define a modified (momentum dependent) four-point function, which is order by order all the diagrams which do not go on-shell below the inelastic threshold.  We depict this modified vertex in Fig.~\ref{fig:vertex}$(a)$.  These $2PI$ diagrams can then be resummed to all orders to produce the scattering matrix, Fig.~\ref{fig:vertex}$(b)$.  It is this resummation that produces the L\"{u}scher relation, relating the finite volume scattering to the infinite volume scattering parameters~\cite{Maiani:1990ca,Hamber:1983vu,Luscher:1986pf,Luscher:1990ux}.  In this way, one can see that the new interactions will not lead to a modification of the structure of the L\"{u}scher relation.

Our last note of caution regards the construction of the interpolating fields.  In the physical basis, the states which become those of definite isospin in the continuum limit are given by Eq.~\eqref{eq:pipiStates}.
However, the interpolating fields are generally constructed with quark fields in the twisted basis, with a known definite \textit{twist} from the physical basis fields.  While this is also true of the $\pi^+\pi^+$ scattering channel, the phase is trivial since there is only one term contributing to the $|2,2\rangle$ state.  Thus if one were to undertake a calculation of these coupled scattering channels, care should be taken in constructing the correct interpolating fields.

\section{Conclusions}
In this report, we have detailed $\pi\pi$ interactions in twisted mass $\chi$PT.  We have shown that through NLO, at maximal twist the corrections to the $I=2, I_3=\pm2$ scattering parameters from discretization errors are identically zero.  However, near maximal twist there are corrections which can modify the expected chiral behavior which we demonstrated by an explicit calculation of the correction to the scattering length.  We found however, that for the dynamical twisted mass lattice configurations which exist today, the expected corrections are negligible.

The $I_3=0$ scattering channels proved to have more significant discretization corrections, most notably a mixing term between the $|2,0\rangle$ and $|0,0\rangle$ states which is relatively large.  In fact, these mixing terms combined with the need for computing quark disconnected diagrams and the expected nonperturbative shift of the twisted mass eigenstates, as discussed in Sec.~\ref{sec:I3=0}, may make a calculation of these $I_3=0$ scattering channels prohibitively complicated.

\acknowledgments
AWL would like to thank Christopher Aubin for useful discussions.  MIB would like to than Paulo Bedaque and Brian Tiburzi for useful discussions.  The work of MIB was supported in part by the
U.S. DOE, Grant No. DE-FG02-93ER-40762.  JWC is supported by the National Science Council and the NCTS of R.O.C..  The work of AWL was supported in part by the U.S. DOE OJI grant DE-FG02-07ER41527.


\end{document}